\UseRawInputEncoding
\documentclass[aps, prl, twocolumn, showpacs,
]{revtex4-1} 
\usepackage{amsmath}
\usepackage{amssymb}
\usepackage{graphicx}
\usepackage{color}

\renewcommand{\vec}[1]{\boldsymbol{#1}}
\begin{document}
	
\title{Resonant four-photon scattering of collinear laser pulses in plasma}
	\author{V.~M.~Malkin}
	\affiliation{Department of Astrophysical Sciences, Princeton University, Princeton, NJ 08540, USA}
	\author{N.~J.~Fisch}
	\affiliation{Department of Astrophysical Sciences, Princeton University, Princeton, NJ 08540, USA}

	\date{\today}

	\begin{abstract}
		
Exact four-photon resonance of collinear planar laser pulses is known to be prohibited by the classical dispersion law of electromagnetic waves in plasma. We show here that the renormalization produced by an arbitrarily small relativistic electron nonlinearity removes this prohibition. The laser frequency shifts in collinear resonant four-photon scattering increase with laser intensities. For laser pulses of frequencies much greater than the electron plasma frequency, the shifts can also be much greater than the plasma frequency and even nearly double the input laser frequency at still small relativistic electron nonlinearities. This may enable broad range tunable lasers of very high frequencies and powers. 
Since the four-photon scattering does not rely on the Langmuir wave, which is very sensitive to plasma homogeneity, such lasers would also be able to operate at much larger plasma inhomogeneities than lasers based on stimulated Raman scattering in plasma.

	\end{abstract}
	\maketitle

\setcounter{secnumdepth}{2}	
\section{Introduction}\label{(Intr)}
Resonant interactions of collinear waves are particularly important because collinear wave packets are less susceptible to transverse slippage and could overlap longer, thus facilitating significant nonlinear transformations  at relatively small intensities.  
However, low order resonant interactions of collinear waves may be forbidden for some classes of wave frequency dependence on wave number,  $\omega(k)$.

In particular, the synchronism conditions for resonant two-into-two wave scattering of collinear waves read
\begin{gather}\label{c1}
k_1+k_2=k_3+k_4\, ,\\ \label{c2}
\omega_1+\omega_2=\omega_3+\omega_4\, , \;\;\; 
\omega_j=\omega(k_j), \;\;\; j=1,2,3,4.
\end{gather}
These reduce, by a simple change of variables
\begin{equation}\label{c3}
k_1=k+q_1\, , \;\; k_2=k-q_1\, , \;\; k_3=k+q_3\, ,\;\; k_4=k-q_3\, ,
\end{equation}
to the equation 
\begin{equation}\label{c4}
f(k,q_1)=f(k,q_3)\, , \;\;\; f(k,q)=\omega(k+q)+\omega(k-q)\, .
\end{equation}
For a monotonic in $q$ function $f(k,q)$, Eq.~(\ref{c4}) has only the trivial solution $q_3=q_1$, for which output waves are identical with input waves, so that there is no ``real" scattering.

In principle, even an arbitrarily small nonlinearity could cause a non-zero shift between wave numbers of output and input waves, thus enabling a real scattering. We will explore this possibility for the relativistic electron nonlinearity of electromagnetic waves in plasma. 

The classical dispersion law for electromagnetic waves of physically ``infinitely small" amplitudes in plasma is
\begin{equation}\label{c5}
\omega(k)=\sqrt{c^2k^2+\omega_e^2}\, , \;\;\;\;  \omega_e=\sqrt{4\pi n_0 e^2/m} \, ,
\end{equation}
where $c$ is the speed of light in vacuum, $m$ is the electron rest mass,  $-e$ is the electron charge, and  $n_0$ is the electron concentration of plasma. 
Under the convention $k>0$, $k_1>k_2$ and $k_3>k_4$, both $q_1$ and $q_3$ are by definition positive. For $q>0$ and $\omega(k) $ given by Eq.~(\ref{c5}),
\begin{gather}\label{c6}
\frac{\partial f(k,q)}{c^2\partial q} = \frac{k+q}{\omega(k+q)} - \frac{k-q}{\omega(k-q)} > 0\, , 
\end{gather}
so that $f(k,q)$ is a monotonically increasing function of $q>0$, which prohibits non-trivial resonant collinear four-photon scattering.  

\begin{figure}[!h]
	\centering
\includegraphics[width=1\linewidth]{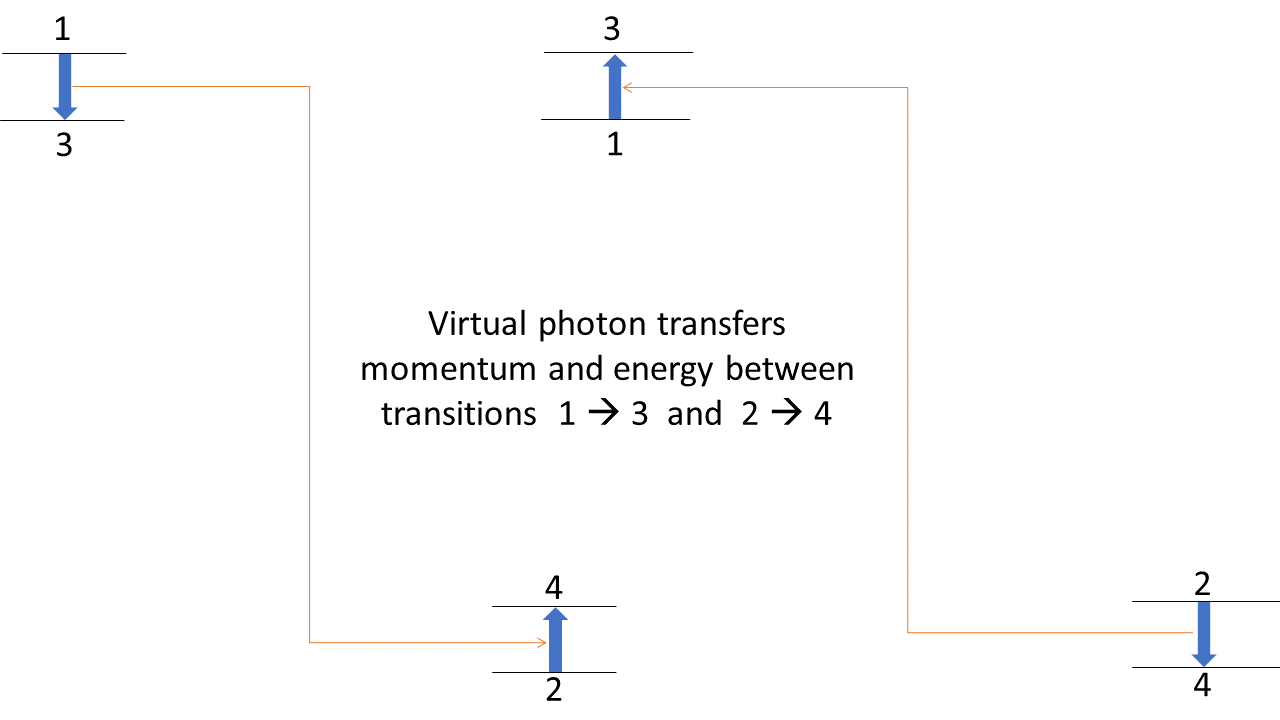}
	\caption{Opposite transitions of the input photons 1 and 2 into the output photons 3 and 4, respectively, coupled via a virtual photon exchange.} 
			\label{fig:2020-4-photon}
\end{figure}

The renormalized resonant process to be explored could remove this prohibition by producing non-zero opposite frequency shifts of pulses 1 and 2. We will conventionally call the scattering ``up-shifting" if it shifts up the higher frequency $\omega_1$, as schematically shown on the right Fig.~\ref{fig:2020-4-photon}, or ``down-shifting" if it shifts down the higher frequency $\omega_1$, as on the left Fig.~\ref{fig:2020-4-photon}.  

The dispersion law Eq.~(\ref{c5}) prohibits also a photon decay into two or three photons. A photon may decay into a photon and electrostatic Langmuir wave of frequency close to the plasma frequency $\omega_e$. This process, called resonant stimulated Raman scattering, received significant attention \cite{1974-PhysFluids-Drake-ParamInstEMWavesPlasma,1975-PhysFuids-Forslund-laser-scattering,1983-PhysFluids-Estabrook-Kruer,1992-PhysFluidsB-McKinstrie-SRFS-RelModul,1992-PRL-Antonsen-Mora,1993-PhysFluidsB-Antonsen-Mora}. We will assume that the Raman scattering is suppressed, which could relatively easily occur due to small inhomogeneities in plasma frequency, detuning the resonance, or due to a longitudinal slippage between the electromagnetic and electrostatic waves having very different group velocities. 
\section{Nonlinear Evolution Equation}\label{NEE}
To facilitate the forthcoming calculations, we will use the nonlinear evolution equation for dimensionless vector-potential  $\vec{a}=e\vec{A}/mc^2$ of electromagnetic field derived in \cite{2020-PRE-Malkin-4photonMJlasers}. This general equation is much simplified in the special case of plane waves propagating along the axis $z$ and polarized along the axis $x$, so that $\vec{a}=a \vec{e_x} $,
\begin{gather}\label{c7}
(\partial^2_{t} - c^2\partial^2_{z} + \omega_e^2){a} =\\
\nonumber
\omega_e^2{a}[1-(\partial^2_{t}+\omega_e^2)^{-1}c^2\partial_{zz}] a^2/2 + O(a^5).  
\end{gather}
Though structurally similar to averaged cubic nonlinear wave equations, Eq.~(\ref{c7}) does not assume any space or time averaging and, therefore, is applicable even for sets of laser pulses of very different wavelengths and frequencies, in contrast to the averaged wave equations used for describing  stimulated Raman scattering of laser pulses in plasma \cite{1974-PhysFluids-Drake-ParamInstEMWavesPlasma,1975-PhysFuids-Forslund-laser-scattering,1983-PhysFluids-Estabrook-Kruer,1992-PhysFluidsB-McKinstrie-SRFS-RelModul,1992-PRL-Antonsen-Mora,1993-PhysFluidsB-Antonsen-Mora}.

For completeness, we give here a simplified derivation of Eq.~(\ref{c7}).  
The derivation starts from the Maxwell equations in Coulomb gauge and Hamilton-Jacobi equation for electron motion in electromagnetic fields  \cite{LandauLifshitz-ClassicalTheoryFields-1975-ch3} (kinetic effects are neglected, because  all the beating phase velocities and electron quiver velocities considered here are much larger than electron thermal velocities):  
\begin{gather}\label{c8}
\vec{H}=\nabla\times\vec{A}, \quad 
\vec{E}=- \partial_{ct}\vec{A}-\nabla\Phi,
\quad  \nabla\cdot\vec{A}=0, \\
\label{c9}
\Box\vec{A} \equiv
\partial^2_{ct}\vec{A} - \Delta\vec{A} = 4\pi\vec{J}/c  - \partial_{ct}\nabla\Phi,\\
\label{c10} 
\vec{J}=-en\vec{P}/\sqrt{m^2+P^2/c^2}, \quad 
\Delta\Phi = 4\pi e (n-n_0), \\
\label{c11} 
\!\!\!\!\!\vec{P}=e\vec{A}/c+\nabla S, \;\; 
\partial_{t} S = e\Phi - c\sqrt{m^2c^2+P^2}+mc^2.\!
\end{gather}
For dimensionless electromagnetic potentials, electron momentum and  action,
\begin{equation}\label{c12}
\!\vec{A} = mc^2\vec{a}/e, \;\, \Phi=mc^2\phi /e, \;\,
\vec{P} = mc\vec{p}, \;\, S=mc^2 s,\!\!\!
\end{equation}
the equations can be reduced to the form
\begin{gather}
\label{c13}
c^2\Box\vec{a} = -\vec{p} (1+p^2)^{-1/2}(\omega_e^2+c^2\Delta\phi)  - c\partial_{t}\nabla\phi,\\
\label{c14} 
\nabla\cdot\vec{a}=0, \;\;
\vec{p}=\vec{a}+c\nabla s, \;\; 
\partial_{t} s = \phi - \sqrt{1+p^2}+1.
\end{gather}
For collinear plane waves, all quantities depend, apart from on time, only on the longitudinal spacial variable $z$. Then, according to the  
equation $\nabla\cdot\vec{a}=0$,  $\vec{a}$ has zero $z$-component, and 
Eqs.~(\ref{c13})-(\ref{c14}) take the form
\begin{gather}
\label{c15} \left[\partial^2_{t} - c^2 \partial^2_{z} + \frac{\omega_e^2+c^2(\partial^2_{z}\phi)}{\sqrt{1 + a^2 + (c\partial_{z} s)^2}} \right] \vec{a} = 0\, ,\\
\label{c16} 
\frac{\omega_e^2+c^2(\partial^2_{z}\phi)}{\sqrt{1 + a^2 + (c\partial_{z} s)^2}}\, c\partial_{z} s  + c\partial_{t}\partial_{z}\phi = 0\, ,\\
\label{c17} \partial_{t} s = \phi - \sqrt{1+a^2 + (c\partial_{z} s)^2}+1\, .
\end{gather}

For mildly relativistic electron quiver velocities in laser pulses, $a\ll 1$, Eqs.~(\ref{c15})-(\ref{c17}) can be expanded in powers of the small parameter $a$. To calculate the 4-photon coupling, the expansion should include terms up to cubic in $a$. 
In a uniform plasma, $\nabla\omega_e=0$,  
with the laser beatings well off the Raman resonance, and leading terms of the electrostatic potential $\phi$ and action $s$ expansions quadratic in $a$ (as will be immediately seen from the result), the expanded Eqs.~(\ref{c15})-(\ref{c17}) take the form
\begin{gather}
\label{c18} \left[\partial^2_{t} - c^2 \partial^2_{z} + \omega_e^2 (1-a^2/2) +c^2(\partial^2_{z}\phi) \right]\vec{a} = O(a^5)\, ,\\
\label{c19} 
\omega_e^2 s +\partial_{t}\phi =O(a^4)\,  ,\\
\label{c20} \partial_{t} s = \phi - a^2/2 + O(a^4)\, .
\end{gather}
Exclusion of $s$ from Eqs.~(\ref{c19})-(\ref{c20}) gives the equation
\begin{equation}\label{c21}
(\partial^2_{t} + \omega_e^2)\phi= a^2\omega_e^2/2 + O(a^4)\, ,
\end{equation}
which can be used now to exclude $\phi$ from Eq.~(\ref{c18}),
\begin{gather}\label{c22}
(\partial^2_{t} - c^2 \partial^2_{z} + \omega_e^2)\vec{a} = \\
\vec{a}[1 - c^2\partial^2_{z}(\partial^2_{t} + \omega_e^2)^{-1} ] a^2\omega_e^2/2  + O(a^5)\, . \nonumber
\end{gather}
For all waves polarized in the same direction, $\vec{a}=a\vec{e_x}$,   Eq.~(\ref{c22}) reduces to  Eq.~(\ref{c7}).
\section{Renormalized dispersion} \label{Disp}
For four-wave sets considered here,
\begin{equation}\label{c23}
a=\sum_{j=1}^{j=4}a_j \exp[\imath(k_{j}z -\omega_j t)] +c.c. + \delta a\, ,
\end{equation}
where $\delta a $ represents small non-resonant beatings generated by nonlinearity. For ``infinitely small" pulses 3 and 4, 
$|a_3|, |a_4|\lll |a_1|, |a_2| $ , 
only pulses 1 and 2 contribute to renormalization of dispersion relations.

Substituting Eq.~(\ref{c23}) in Eq.~(\ref{c7}) and collecting all terms varying like $\exp[\imath(k_{j}z -\omega_j t)]$ lead to the following renormalized dispersion relations:
\begin{gather}
F_{j,l}= 3 - \frac{c^2(k_j-k_l)^2}{(\omega_j-\omega_l)^2-\omega_e^2} -
\frac{c^2(k_j+k_l)^2}{(\omega_j+\omega_l)^2-\omega_e^2}\, , \label{c24} \\
\nonumber
c^2k_1^2+\omega_e^2-\omega_1^2 \approx \omega_e^2\, 
\left(|a_1|^2 F_{1,1}/2   +  |a_2|^2 F_{1,2}\right)\, ,\\
 \nonumber
c^2k_2^2+\omega_e^2-\omega_2^2 \approx \omega_e^2\,
\left(|a_2|^2 F_{2,2}/2   +  |a_1|^2 F_{1,2}\right)\, , \\
\nonumber
c^2k_3^2+\omega_e^2-\omega_3^2 \approx \omega_e^2\, 
\left(|a_1|^2 F_{1,3}   +  |a_2|^2 F_{2,3}\right)\, ,\;\;\;\\
\label{c25} 
\;\;\; c^2k_4^2+\omega_e^2-\omega_4^2 \approx \omega_e^2\, 
\left(|a_1|^2 F_{1,4}   +  |a_2|^2 F_{2,4}\right)\, . 
\end{gather}
These relations can be simplified for laser frequencies significantly exceeding the plasma frequency.
The terms having sums of laser frequencies in denominators can be approximately replaced then by their vacuum value 1. Neglecting  corrections containing extra small factors of order of $\omega_e^2/\omega_j^2$, Eqs.~(\ref{c24})-(\ref{c25}) reduce to:
\begin{gather}
F_{j,l}\approx 2 - \frac{c^2(k_j-k_l)^2}{(\omega_j-\omega_l)^2-\omega_e^2}
\label{c26} \\\nonumber
\omega_1\approx ck_1 + \frac{\omega_e^2}{2\omega_1} 
\left(1 - |a_1|^2  - |a_2|^2 F_{1,2} \right) ,\\
 \nonumber
\omega_2\approx ck_2 + \frac{\omega_e^2}{2\omega_2} 
\left(1 - |a_2|^2  - |a_1|^2 F_{1,2} \right) ,\\
\nonumber
\omega_3\approx ck_3 + \frac{\omega_e^2}{2\omega_3}
\left(1 - |a_1|^2F_{1,3} - |a_2|^2F_{2,3} \right)\, , \\
\label{c27} 
\;\;\;\;\;\;\omega_4\approx ck_4 + \frac{\omega_e^2}{2\omega_4}
\left(1 - |a_1|^2F_{1,4} - |a_2|^2F_{2,4}\right)\, . 
\end{gather}

\section{Four-photon resonance}\label{Res}
The renormalized dispersion law Eqs.~(\ref{c26})-(\ref{c27}) gives the following frequency mismatch in the temporal synchronism condition for four-photon scattering:
\begin{gather}\label{c28}
\delta\omega\equiv\omega_1+\omega_2 - \omega_3-\omega_4\approx \\ \nonumber
\frac{\omega_e^2k (q_1^2-q_3^2)}{ck_1k_2k_3k_4}(1-|a_1|^2 - |a_2|^2) + \\ \nonumber  
\frac{\omega_e^2(k_1|a_1|^2+k_2|a_2|^2)}{2ck_1k_2} \left(1-F_{1,2}\right) - \\ \nonumber  
\frac{\omega_e^2|a_1|^2}{2ck_3k_4}\left[k_3(1-F_{1,4}) +k_4(1-F_{1,3})
\right] - \\ \nonumber  
\frac{\omega_e^2|a_2|^2}{2ck_3k_4}\left[k_3(1-F_{2,4}) +k_4(1-F_{2,3}) 
\right]\, .
\end{gather}
The  four-photon resonance condition, $\delta\omega=0 $, at which $F_{1,3}=F_{2,4}$ and $F_{2,3}=F_{1,4}$, can be  presented in the form 
\begin{gather}\nonumber
2k(q_1^2-q_3^2)\approx  
k_3k_4(k_1|a_1|^2+k_2|a_2|^2)(F_{1,2}-1) +  
 \\ \nonumber  
k_1k_2(k_4|a_1|^2+k_3|a_2|^2)(1-F_{1,3}) +  \\ 
k_1k_2(k_3|a_1|^2+k_4|a_2|^2)(1-F_{1,4})\, . \label{c29}
\end{gather}
For $|a_2|\approx |a_1|$, it simplifies to
\begin{gather}\label{c30}
(q_1^2-q_3^2)/|a_1|^2\approx  k_3k_4  (F_{1,2}-1) - k_1k_2F\, , \\ \label{c33}  
F=F_{1,3}+F_{1,4}-2\, . 
\end{gather}
We will use this condition below to calculate the resonant manifold in the space of parameters $k$, $q_1$, $q_3$, $k_e\equiv \omega_e/c$ and $|a_1|$. But first, we will derive a general formula for the rate of the resonant four-photon scattering, in order to be able to calculate this rate right away in each regime.
\section{Four-photon scattering rate}\label{rate}
The dominant terms producing scattering via the above four-photon resonance in Eq.~(\ref{c7}) give the following evolution equations for very small slowly varying amplitudes of scattered pulses $3$ and $4$:
\begin{gather}\label{c31}
2\imath \omega_3 \partial_t a_3 \approx  a_1a_2a_4^*\omega_e^2F, \\
\label{c32}
2\imath \omega_4 \partial_t a_4 \approx a_1a_2a_3^*\omega_e^2 F\, . 
\end{gather}
In the growing mode, the phases are synchronized as
\begin{equation}\label{c34}
\arg{a_1} + \arg{a_2}+\arg{F} =\arg{a_3} + \arg{a_4}+\pi/2 ,
\end{equation}
and
\begin{gather}\label{c35}
|a_3|\sqrt{\omega_3}\approx |a_4|\sqrt{\omega_4} \propto \exp{\gamma t},\\
\label{c36}
\gamma \approx \frac{\omega_e^{2}\,|Fa_1a_2|}{2\sqrt{\omega_3\omega_4}} .
\end{gather}

\section{Small resonant shift $|q_3-q_1|\ll k_e $} \label{SmallShift}

First, consider the resonance condition Eq.~(\ref{c30}) in the limit $|a_1|\rightarrow 0$.
Well off the Raman resonances, corresponding to zeros of $F_{1,j}$ denominators on the right hand side, it implies $q_3\rightarrow q_1 $. Then, $k_3\rightarrow k_1$, $k_4\rightarrow k_2$, $F_{1,3}
\rightarrow 2$, $F_{1,4}\rightarrow F_{1,2}$ and Eq.~(\ref{c30})
reduces to
\begin{equation}\label{c37}
q_3^2- q_1^2\approx|a_1|^2\, k_1k_2\, .
\end{equation}
As seen,  $q_3>q_1$, so that the scattering up-shifts the photon 1, like in the scheme on the right Fig.~\ref{fig:2020-4-photon}. 
The scattering rate Eq.~(\ref{c36}), with $F$ from Eqs.~(\ref{c33}) and (\ref{c26}), in resonance Eq.~(\ref{c37}), is  
\begin{equation}\label{c38}
\gamma \approx \frac{\omega_e^{2}\,|a_1|^2}{2\sqrt{\omega_1\omega_2}} 
\left| \frac{(q_1+q_3)^2-2k_e^2}{(q_1+q_3)^2-k_e^2}\right| .
\end{equation}
Thus, even an arbitrarily small relativistic electron nonlinearity  enables non-trivial resonant collinear four-photon scattering with a non-zero frequency up-shift. 

For regimes which are well off the Raman resonances at $q_3+q_1\approx k_e$ and $q_1\approx k_e/2$, formulas Eqs.~(\ref{c37})-(\ref{c38}) remain applicable as long as $q_3-q_1\ll k_e$, namely, at 
\begin{equation}\label{c39}
|a_1|^2\ll  \frac{k_e^2+2k_eq_1}{k_1k_2}\, .
\end{equation}

\section{Large resonant shift $|q_3-q_1|\gg k_e $} \label{LargeShift}

For $ |q_3-q_1|\gg k_e$, Eqs.~(\ref{c30}) and (\ref{c36}), supplemented by (\ref{c26}) and (\ref{c33}),  approximately reduce to
\begin{gather}\label{c40}
\frac{q_1^2-q_3^2}{k_e^2|a_1|^2}\approx     
\frac{2k_1k_2(q_1^2+q_3^2)} {(q_1^2-q_3^2)^2} - \frac{k^2-q_3^2} {4q_1^2- k_e^2}  , \\
\label{c41}
\gamma \approx \frac{ck_e^{4}\,|a_1|^2(q_1^2+q_3^2)} {(q_1^2-q_3^2)^2\,\sqrt{k^2-q_3^2}}\, .
\end{gather}
Eq.~(\ref{c40}) can be rewritten in the form 
\begin{gather}\label{e42}
Y\approx \frac{X^3}{X^2-2QX-4Q}\, ,\;\;\;\; Q=4-\frac{k_e^2}{q_1^2}\, ,\\
\label{e43}
X= \frac{q_3^2}{q_1^2}-1\, ,\;\;\;\;   
Y= \frac{k_1k_2k_e^2|a_1|^2}{q_1^2(q_1^2Q+k_e^2|a_1|^2)}\, .
\end{gather}

\begin{figure}[h!]
	\centering
	\includegraphics[width=1\linewidth]{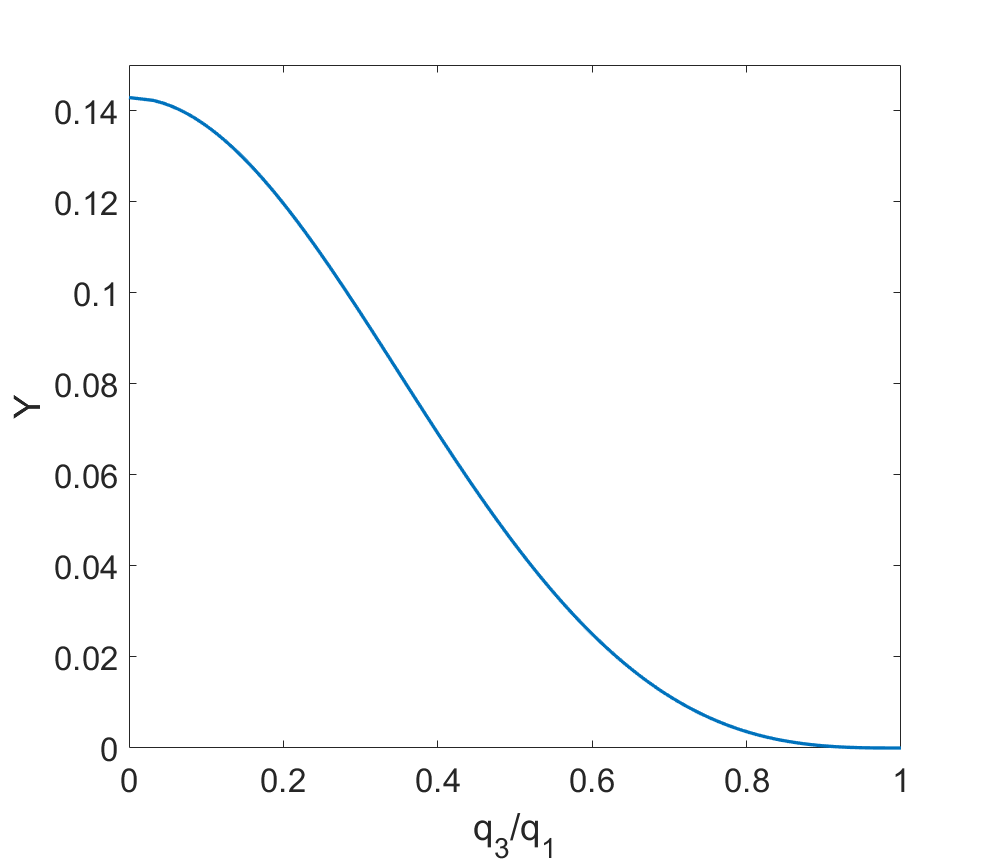}
	\caption{The resonant manifold Eq.~(\ref{d44}) for down-shifting regimes $q_1-q_3\gg k_e$.}
	\label{fig:2020-4-photon-f2}
\end{figure}

For the down-shifting regimes, $q_1-q_3\gg k_e \Longrightarrow Q\approx 4$, Eqs.~(\ref{e42})-(\ref{e43}) approximately reduce to
\begin{equation}\label{d44}
\frac{k_1k_2k_e^2|a_1|^2}{4q_1^4}\approx Y\approx \frac{X^3}{X^2-8X-16}\, ,\;\;\; X= \frac{q_3^2}{q_1^2}-1\, .
\end{equation}
This resonant manifold is shown in the Fig.~\ref{fig:2020-4-photon-f2}.
As seen, $q_3/q_1$ increasing from 0 to 1 corresponds to $Y$ monotonically decreasing from $1/7$ to 0. The parameters need to satisfy the requirement 
\begin{equation}\label{e44}
1\gg |a_1|^2\approx \frac{4q_1^4Y}{k_1k_2k_e^2}\, .
\end{equation}

According to Eq.~(\ref{d44}), at $k_e/q_1\ll 1-q_3/q_1\ll 1$  
$$\frac{X}{2}\approx -\frac{q_1-q_3}{q_1}\, , \;\;\; 
\frac{k_1k_2k_e^2|a_1|^2}{4q_1^4}\approx Y\approx \frac{(q_1-q_3)^3}{2q_1^3}  \ll 1\, .$$ 
The down-shift is indeed much greater than the Raman scattering down-shift, $q_1-q_3\gg k_e$, at 
\begin{equation}\label{c44}
|a_1|^2\gg \frac{2k_eq_1}{k_1k_2}\, . 
\end{equation}
The growth rate Eq.~(\ref{c41}) at $1-q_3/q_1\ll 1$ is
\begin{gather}\label{c42}
\gamma \approx \frac{ck_e^{8/3}q_1^{2/3}|a_1|^{2/3}} {2^{1/3}(k_1k_2)^{7/6}}\, .
\end{gather}

As was schematically shown on the left in Fig.~\ref{fig:2020-4-photon}, the down-shifting four-photon scattering ($q_3<q_1$) makes photon wave-numbers closer to the average wave-number $k$. 
This tendency might be viewed as a dynamic counterpart of the tendency to Bose-Einstein condensation in kinetic regimes of four-wave scattering \cite{1992-ZakharovLvovFalkovich-KolmogorovTurb, 1996-PRL-Malkin, 2015-PRE-Falkovich,2018-PRE-Malkin-OpticalTurb}. The caveat is that all waves and beatings involved here are of the same nature. Otherwise, the tendency may change to up-shifting, like in the regime Eq.~(\ref{c37}) where the four-photon scattering is noticeably mediated by quasi-electrostatic beatings. 

\begin{figure}[h!]
	\centering
	\includegraphics[width=1\linewidth]{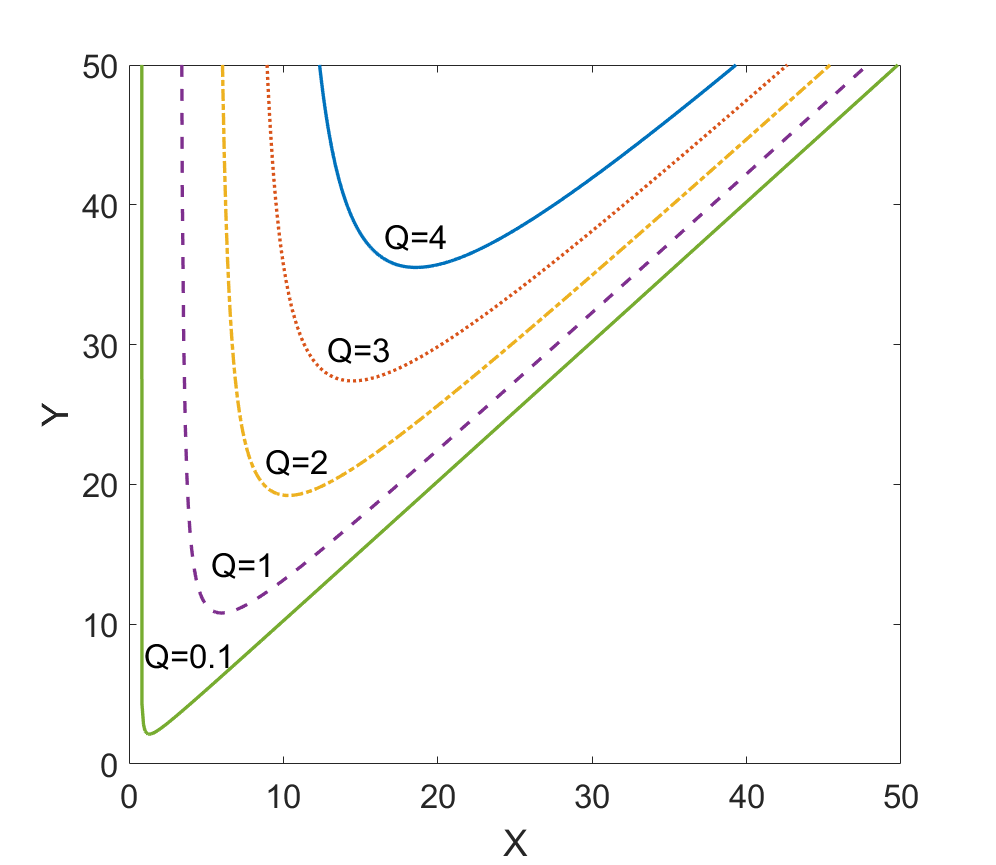}
	\caption{The resonant manifold Eq.~(\ref{e42}) for up-shifting regimes $q_3-q_1\gg k_e$.}
	\label{fig:2020-4-photon-f3}
\end{figure}

For the up-shifting regimes, $q_3-q_1\gg k_e $, 
the resonant manifold given by Eqs.~(\ref{e42})-(\ref{e43}) is located at $Q>0$, $X>Q+\sqrt{Q^2+4Q}\,$.
It is shown in the Fig.~\ref{fig:2020-4-photon-f3}. At $X\gg 2+2Q$, Eq.~(\ref{e42}) approximately reduces to $Y\approx X+2Q$. 

Eq.~(\ref{e43}) can be rewritten in the form 
\begin{equation}\label{c48}
Y=\frac{k_1k_2}{q_1^2(Z^{-1}+1)}\, , \;\;\;\; Z=\frac{k_e^2|a_1|^2}{q_1^2Q}>0\, .
\end{equation}
For $X\gg 2+2Q$, when $Y\approx X\approx q_3^2/q_1^2\gg 1$, it follows
\begin{equation}\label{c51}
q_3\approx k/\sqrt{1+Z^{-1}}\, .
\end{equation}
The parameters need to satisfy the requirement 
\begin{equation}\label{c49}
1\gg |a_1|^2= (4q_1^2/k_e^2-1)Z\,    .
\end{equation}
The allowed $Z\ll (4q_1^2/k_e^2-1)^{-1}$ is small, $Z\ll 1$, at $4q_1^2/k_e^2-1\gtrsim 1$, but can be larger, $Z\gtrsim 1$, at 
$0<4q_1^2/k_e^2-1\ll 1$. 
To stay well off the Raman resonance at $2q_1\approx k_e$, having the width $|2q_1-k_e|\sim k_e^2|a_1|/2k$ in the regime $|a_1|\gg k_e/k$ of strongly coupled Stokes and anti-Stokes waves  \cite{1983-PhysFluids-Estabrook-Kruer, 1992-PhysFluidsB-McKinstrie-SRFS-RelModul}, it should be $ Z^{-1}\gg k_e/k|a_1|  $. This can be also put in the form $k_4=k-q_3\approx kZ^{-1}/2\gg k_e/2|a_1|$, or 
\begin{equation}\label{d51}
|a_1|\gg k_e/2k_4\, .
\end{equation}

The four-photon scattering rate Eq.~(\ref{c41}) in the regime Eq.~(\ref{c51}) is 
\begin{equation}\label{48}
\gamma \approx \frac{ck_e^{4}|a_1|^2(Z+1)^{3/2}}{k^3Z}\, .
\end{equation}
For  $Z\ll 1$, it reduces to
\begin{equation}\label{49}
\gamma\approx \frac{ck_e^{4}|a_1|^2}{k^3Z} \approx \frac{ck_e^{2}(4q_1^2-k_e^{2})} {k^3}
\end{equation}
and does not depend on the laser intensity. This behavior departs from that of the common four-wave scattering rate proportional to the wave intensity. It also departs from behavior of the modified rate Eq.~(\ref{c42}) proportional to the cubic root of intensity as common for strongly coupled four-wave scattering regimes. The rate  Eq.~(\ref{49})  contains the small factor $(4q_1^2-k_e^{2})/k^2 $ instead of the common small factor proportional to intensity (apart from the factor $\omega_e^2/\omega $ common for all regimes).

For $1\ll Z\ll k|a_1|/k_e$, when the resonant four-photon scattering nearly doubles the laser frequency $k_3\equiv k+q_3 \approx 2k-k/2Z$, the rate Eq.~(\ref{48}) reduces to
\begin{equation}\label{50}
\gamma \approx \frac{ck_e^{4}|a_1|^2\sqrt{Z}}{k^3}\approx \frac{ck_e^{5}|a_1|^3}{k^3\sqrt{4q_1^2-k_e^{2}}}\, .
\end{equation}
In contrast to the case Eq.~(\ref{49}), this rate depends on laser intensity even more strongly than does the common four-photon scattering rate.

The rate Eq.~(\ref{48}), as a function of $Z$, is minimal at $Z=2$, where  
\begin{equation}\label{51}
\gamma \approx \frac{3\sqrt{3}\, ck_e^{4}|a_1|^2}{2k^3}\, ,
\;\;\;\; k_3\approx 1.8 k\, .
\end{equation}
Even around this minimum, the rate can be sufficient to accomplish the scattering within small enough propagation distances.
 
For example, at $Z=1$, $k\approx 20\, k_e$ and $|a_1|^2\approx 0.1$, the resonant four-photon scattering produces the output laser pulse 3 of 1.7  the input laser frequency. The small-amplitude pulse 3 grows exponentially 
with the propagation distance and one exponentiation occurs within the  length $c/\gamma\approx  9\times 10^{4}\,\lambda $. 
For the input laser wavelength of, say, $ \lambda = 1/3\,\mu$m, this length is $c/\gamma\approx 3$~cm.  
	
\section{Summary}
A new class of basic nonlinear processes has been identified, whereby collinear laser pulses can undergo an exactly resonant four-photon scattering, prohibited according to the classical dispersion law. The resonance is made possible by the intensity-dependent nonlinear frequency renormalization of the laser pulses. Despite being a higher order nonlinear process than the three-wave process of stimulated Raman scattering, the resonant renormalized four-photon scattering can occur in a relatively small propagation distance, so as to be of experimental interest in laboratory settings. Moreover, it can have important advantages over the stimulated Raman scattering: 
\begin{itemize}
\item Because it does not rely on the Langmuir wave, which is very sensitive to plasma homogeneity, the four-photon scattering can tolerate much larger plasma inhomogeneities than can stimulated Raman scattering.
\item Due to the smallness of longitudinal slippage between collinear laser pulses at large laser-to-plasma frequency ratios, the four-photon scattering can normally operate even for much shorter laser pulses than can stimulated Raman scattering.
\item Frequency shifts produced by the resonant renormalized four-photon scattering can be tuned over a broad range by varying the intensity of the pulses, while the stimulated Raman scattering can shift the laser frequency only by the electron plasma frequency.
\item For laser frequencies much greater than the electron plasma frequency, frequency shifts produced by the resonant renormalized four-photon scattering frequency can also be much greater than the electron plasma frequency at still just a mildly relativistic electron nonlinearity.
\item The frequency shift can in fact be almost as large as the laser frequency, leading to output frequencies nearly twice the input frequency. The large frequency upshifts may enable an all-optical resonant frequency multiplication cascade in collinear geometry, which would be free of challenges associated with the transverse slippage of non-collinear laser pulses \cite{2020-PRE-Malkin-4photonMJlasers}. 
\end{itemize}
\section{Acknowledgment}
The work is supported by Grants NNSA DENA0003871 and AFOSR  FA9550-15-1-0391.
%

\end{document}